\documentclass[12pt]{article}

\begin{document}
\vspace*{-.6in} \thispagestyle{empty}
\begin{flushright}
CALT-68-2318\\ CITUSC/01-002
\end{flushright}
\baselineskip = 20pt

\vspace{.5in}
{\Large
\begin{center}
Anomaly Analysis of Brane-Antibrane Systems
\end{center}}

\begin{center}
John H. Schwarz and Edward Witten\\
\emph{California Institute of Technology, Pasadena, CA  91125, USA\\
and\\
Caltech-USC Center for Theoretical Physics\\
University of Southern California, Los Angeles, CA 90089, USA}
\end{center}
\vspace{1in}

\begin{center}
\textbf{Abstract}

\end{center}
\begin{quotation}
\noindent

String theories with branes can often be generalized by adding
brane-antibrane pairs.  We explore the cancellation of anomalies in this
more general context, extending the familiar anomaly-cancelling mechanisms,
both for ten-dimensional string theories with D-branes and for certain
supersymmetric compactifications.

\end{quotation}

\newpage


\section{Introduction}

Analysis of spacetime anomalies has played an important
role in our understanding of string theories and branes.
String theories with branes can be generalized by adding
brane-antibrane pairs, including spacetime filling branes as in
\cite{Srednicki:1998mq}.  In the present paper, we will analyze
anomaly cancellation in this more general context, showing how
the familiar mechanisms for anomaly cancellation can be generalized.
We consider the Type IIB theory with D9-D$\bar 9$ pairs in section 2,
the Type I theory with such pairs in section 3, and a nonsupersymmetric
but tachyon-free ten-dimensional string theory that has D-branes
in section 4.  In sections 5 and 6 we extend the analysis to consider
Type I compactifications on smooth K3 surfaces and on a simple K3 orbifold.
In these examples, we consider D5-D$\bar 5$ as well as D9-D$\bar 9$ pairs.
\def\Tr{{\rm Tr}}

\def\tilde{\widetilde}
\section{Type IIB Theory}

We begin by considering the Type IIB theory with added D9-branes
and anti-D9-branes. This system was considered first in
\cite{Srednicki:1998mq}. In this section we will review the
analysis of that paper filling in some of the details. Our reason
for doing this is that we want to start with the simplest example
of a chiral theory with added spacetime-filling brane-antibrane
pairs in order to set the stage for the more complicated theories
that we will discuss later.  The anomaly cancellation mechanism
that we will describe is closely related to the analysis
considered in \cite{Green:1997dd}   to determine certain effective
couplings on branes. The main difference is that we focus on D9-
and D$\bar 9 $-branes, while D$p$-branes of $p<9$ were the focus
of that paper.  Type II anomaly cancellation including global
anomalies has been analyzed in a somewhat abstract setting in
\cite{Freed:2000tt}; some illustrative cases of the global
anomalies were studied in \cite{Moore:2000gb}.

Cancellation of Ramond-Ramond (R-R) tadpoles requires that the total D9-brane
charge should vanish. Therefore one must require that an equal
number $n$ of D9-branes and anti-D9-branes are present.

\subsection{The Spectrum}

The $n$ D9-branes have associated massless fields, arising as
excitations of 99 open-strings, that correspond to a
supersymmetric $U(n)$ gauge theory. The adjoint Majorana--Weyl
fermions have the same chirality (call it left-handed) as the
gravitinos of the supergravity multiplet. The adjoint
representation decomposes into irreducible representations with
dimensions $n^2 -1$ and $1$. The singlet fermion can be identified
as the Goldstino associated to the supersymmetry broken by the
presence of the branes. (See \cite{Yoneya:2000qe} for the
interpretation of the singlet massless fermion on a brane as a
Goldstone fermion of spontaneously broken supersymmetry.)
Similarly, the anti-D9-branes carry a second supersymmetric $U(n)$
gauge theory. Its fermions have the same chirality (left-handed)
as those of the branes. The singlet fermion of the anti-branes is
the Goldstino associated to breaking of the other supersymmetry of
the IIB theory. So, when both the branes and anti-branes are
present, the supersymmetry is completely broken. The fact that
they each carry fermions of the same chirality is clearly
required, since the two supersymmetries of the IIB theory have the
same chirality. Altogether, the combined system is a
nonsupersymmetric theory with $U(n) \times U(n)$ gauge symmetry.

Such a system is unstable, of course, since the branes and
anti-branes will tend to annihilate. This is reflected in the
presence of tachyons in the spectrum. Specifically, the oriented
Type IIB open strings that connect the branes to the anti-branes
have the opposite GSO projection from those that connect branes to
branes or anti-branes to anti-branes. As a result, the spectrum
contains tachyons $T$ in the bifundamental representation $(n,
\bar n)$. These scalar fields are complex, of course. In addition
to the tachyons, the open strings connecting branes to anti-branes
also give massless fermions in the bifundamental representation.
These fermions have the opposite chirality (right-handed) to the
other ones as a consequence of the opposite GSO projection.

\subsection{Anomaly Analysis}

Even though the brane-antibrane system in question is unstable, it
should make sense within the context of perturbation theory. The
point where the tachyon field vanishes ($T=0$) corresponds to a
local maximum of the tachyon potential, and thus it is part of a
classical solution.  The one-loop effective action in an expansion
around this solution should be well-defined, even though the
solution is unstable, and in particular it should have a
well-defined phase. Therefore, the various gauge and gravitational
anomalies, which arise as one-loop effects, should cancel. The
chiral fields of the Type IIB supergravity multiplet give
cancelling contributions to the gravitational anomalies, just as
they do in the absence of the D-branes
\cite{Alvarez-Gaume:1984ig}. However, there are now additional
chiral fermions in the spectrum, which also contribute to
anomalies. So that needs to be analyzed. We will begin by setting
$T=0$, but later we will incorporate the dependence on $T$.

Let us use subscripts 1 and 2 to refer to the two $U(n)$ groups.
Then the anomaly contributed by the chiral fermions associated to
the branes is characterized by the 12-form part of the following
expression:
\begin{equation} \label{anomaly}
I = \Big(  {\rm Tr} e^{iF_1} + {\rm Tr} e^{iF_2} - {\rm tr}
e^{iF_1} \, {\rm tr} e^{-iF_2} - {\rm tr} e^{-iF_1} \, {\rm tr}
e^{iF_2}\Big) \hat A(R)
\end{equation}
Here $\hat A(R)$ is the Dirac index. The symbol Tr refers to the
adjoint representation, whereas the symbol tr refers to the
fundamental representation. The coefficients $+ 1$ are introduced
for the left-handed Majorana--Weyl fermions, whereas the
coefficient $-1$ appears for right-handed Weyl fermions and their complex
conjugates.

The adjoint representation of $U(n)$ is given by the product $n
\times \bar n$. As a result, using a basic property of the Chern
character, we have
\begin{equation}
{\rm Tr} e^{iF} = {\rm tr} e^{iF} {\rm tr} e^{-iF} .
\end{equation}
This enables us to recast equation (\ref{anomaly}) in the form
$I = Y \bar Y$, where
\begin{equation}
\label{Yeqn} Y = \Big({\rm tr} e^{iF_1} - {\rm tr} e^{iF_2}\Big)
\sqrt{\hat A(R)}.
\end{equation}
and
\begin{equation}
\bar Y = \Big({\rm tr} e^{-iF_1} - {\rm tr} e^{-iF_2}\Big)
\sqrt{\hat A(R)}.
\end{equation}
The anomaly polynomial $I_{12}$ is not zero, but because of the
factorization $I=Y\bar Y$, the anomalies can be
cancelled by the addition of suitable Chern--Simons counterterms
\cite{Green:1984sg}.

The determination of the anomaly cancelling counterterms has some
(inconsequential) ambiguities, which we will resolve by making the
most symmetrical choice. This  will give the most elegant
formulas, and it will make contact with previous results in the
literature. Now let us define $\Omega$ by $Y = d \Omega$, where it
is understood that Chern-Simons terms are introduced for the gauge
field factor leaving the factor $\sqrt{\hat A(R)}$ in tact. This
is desirable because $\sqrt{\hat A(R)}$ has a constant term,
whereas the constant terms arising from the gauge field factor
cancel. Let us denote the variation of $\Omega$ under local gauge
transformations by $\delta \Omega = d \Lambda$.  The anomaly can
now be cancelled by introducing the nontrivial gauge
transformation rules for the R-R fields $C = -i C_0 + C_2 + i C_4
- C_6 -i C_8$
\begin{equation}
\delta C = \Lambda.
\end{equation}
It follows that the gauge-invariant R-R field strengths are given
by $H = dC - \Omega$. In particular, this implies that the R-R
scalar $C_0$ is eaten by the relative $U(1)$ gauge field ${\rm tr}
A_1 - {\rm tr} A_2$, which then becomes massive. We also note the
Bianchi identity $dH + Y = 0$.

Now we can write down the anomaly-cancelling Chern--Simons term in
the form
\begin{equation} \label{CSaction}
S_{\rm CS} = \mu \int C Y \, ,
\end{equation}
where $\mu$ is a normalization constant. This expression is real
with the phase choices made in the definition of $C$. The fact that
R-R fields other than $C_2$ can be involved in anomaly cancellation
was first recognized in \cite{Sagnotti:1992qw}. An interpretation of the
Chern--Simons term in terms of anomaly inflow was given
in \cite{Green:1997dd}.

In writing the couplings, we have ignored the self-duality of the
R-R fields.  One standard approach to dealing with this
is to treat half of the $C$'s as independent fields, in which case
half of the anomaly-canceling couplings remain as we have written
them and the other half become contributions to the Bianchi
identities.  The self-duality of the ``middle'' R-R field
$G_5=dC_4$ makes this procedure subtle to interpret for Type IIB.
There are various procedures for dealing with this, and we will
not discuss the issue here. A more abstract way of interpreting
actions with self-dual R-R fields is in \cite{Freed:2000tt}.

\subsection{Inclusion of the Tachyon Field}

A natural generalization of the Chern--Simons action to include the
dependence on the bifundamental tachyon fields $T$ has been
obtained by other authors 
\cite{Kennedy:1999nn} \cite{Kraus:2000nj} \cite{Takayanagi:2000rz}. 
The result can be
elegantly described as replacing the factor $\Big( {\rm tr}
e^{iF_1} - {\rm tr} e^{iF_2}\Big)$, which appears in the
expression (\ref{Yeqn}) for $Y$, by ${\rm Str} e^{i {\cal F}}$,
where ${\cal F}$ is the curvature of an object that has been
called the superconnection
\cite{Quillen:1985}.  The curvature of the superconnection is
given by
the $2n \times 2n$ matrix
\begin{equation}
i{\cal F}  = \left( \begin{array}{cc}i F_1 - T  T^{\dagger} & D
T^{\dagger}
\\ DT &  i F_2 - T^{\dagger} T \end{array} \right) .
\end{equation}
The supertrace Str is the difference of the traces of the upper
left block and the lower right block, as usual. Clearly, ${\rm
Str} e^{i {\cal F}}$ reduces to our previous expression for $T=0$.
In fact, this notation is convenient even in that case.

If the tachyon field can be considered large, then ${\rm
Str}e^{i{\cal F}}$ is very small except near zeroes of $T$ -- or
at least points at which some eigenvalues of $T$ are zero. Indeed,
in this situation the diagonal elements of $e^{i{\cal F}}$ are
proportional to $e^{-TT^\dagger}$ and $e^{-T^\dagger T}$, which
can be considered small.  Note that in the vacuum, $T$ is a
unitary matrix times a fixed constant, so $TT^\dagger $ and
$T^\dagger T$ are equal and are multiples of the identity.  So if
we can ignore $F_1$, $F_2$, and $DT$, then ${\rm Str}e^{i{\cal
F}}$ vanishes pointwise.  In practice, an important case is that
on the complement of a submanifold that is interpreted as a
$p$-brane world-volume for some $p<9$, $T$ has its vacuum
expectation value (up to a gauge transformation).  Such a tachyon
field describes \cite{Sen:1998sm} D9-D$\bar 9$ annihilation to a
D$p$-brane. If all length scales in the problem are very large
compared to the string scale,  then in the usual generalized
``vortex'' configuration describing tachyon condensation, $F_1$,
$F_2$, and $DT$ all vanish exponentially fast with the distance
from the D$p$-brane. So in this limit, the anomalous couplings are
given by a differential form that  has its support on the
D$p$-brane, rather than on the 9-branes where we started.

Superconnections were introduced in the first place
\cite{Quillen:1985} precisely to give such an analytic proof of
``localization'' of various topological quantities (such as the
anomalies of interest to us here) as well as to explain various
physical results. From the point of view of the application to
D9-D$\bar 9$ annihilation, the fact that this works out correctly
is one consistency  check on the claim that a D9-D$\bar 9$ system
with no net D9-brane charge and  carrying suitable gauge fields
can annihilate to a D$p$-brane with $p<9$.  This check of the
tachyon condensation story is not really independent of analyses
in the previous literature, but it is perhaps an interesting way
to look at things.

\section{Type I Theory}

We now wish to repeat the analysis of the preceding section for
the Type I theory. Recall that this theory has an orientifold
plane carrying $-32$ units of R-R charge, which is cancelled by
the contribution of 32 D9-branes. The D9-branes give $SO(32)$
gauge symmetry. They do not break any supersymmetry, and
accordingly the open string spectrum contains no massless
Goldstinos. The various local anomalies cancel as a result of a
well-known analysis analogous to that of the preceding section
\cite{Green:1984sg}. We now wish to generalize this setup to
include $n$ additional D9-brane anti-D9-brane pairs. This system
has been considered previously by Sugimoto in
\cite{Sugimoto:1999tx}. Our purpose in reviewing these results is
to fill in some details and to set the stage for more complicated
examples.

\subsection{The Spectrum}

The gauge group of the $32+n$ D9-branes is expected to be
$SO(32+n)$, and that of the $n$ anti-D9-branes is expected to be
$SO(n)$. Given our knowledge of the $n=0$ case, it is pretty clear
that the D9-branes should give a supersymmetric $SO(32+n)$ gauge
theory. Thus the 99 open strings should give massless gauge bosons
and massless left-handed MW fermions, each in the adjoint
representation. As before, the $9\bar9$ open strings have the
opposite GSO projection. Therefore they give real tachyon fields
in the bifundamental representation $(32+n, n)$ and right-handed
MW fermions that are also in the bifundamental representation
$(32+n, n)$.

This still leaves the question of the spectrum of $\bar 9 \bar 9$
open strings.  Momentarily we will sketch how to deduce this
spectrum from first principles. For the moment, we simply look for
a choice that satisfies some reasonable expectations. In this
problem, we do not expect any tachyons, but we do expect massless
vector fields in the adjoint of $SO(n)$. We expect the
anti-D9-branes to break the supersymmetry, and therefore there
should be a massless left-handed MW Goldstino, which is a singlet
of the gauge group. Since there is no singlet gauge boson this
tells us that the anti-D-brane spectrum cannot be supersymmetric.
That is okay -- there is no reason that it should be. To figure
out what other fermions are required, let us consider the minimal
requirement of anomaly cancellation -- namely, the cancellation of
the ${\rm tr} R^6$ term. This tells us that the branes and the
anti-branes should contribute a net of 496 left-handed MW
fermions. We already have $(32+n)(31+n)/2$ from the 99 open
strings. From this we must subtract $n(32+n)$ to take account of
the right-handed MW fermions associated to the $9\bar9$ open
strings. Therefore, to end up with 496, the $\bar 9 \bar 9$ open
strings need to contribute a net of $n(n+1)/2$ left-handed MW
fermions. The obvious way to achieve this is with the sum of a
symmetric traceless tensor and a singlet.\footnote{This result is
related by T duality to an analogous result for D2-branes obtained
in \cite{Berkooz:1999ji}.} It is very gratifying that there is a
singlet, since it is the required Goldstino.

\subsection{Microscopic Derivation}

Now we will give a microscopic derivation of the spectrum.

One way to describe a system with D-antibranes as well as D-branes
is to assign variable statistics to the Chan-Paton states
\cite{Witten:1998cd}.  Let us briefly describe how this works.
Consider a system of D-branes with a Chan-Paton label that takes
$p+q$ values; take the first $p$ states to be bosonic and the last
$q$ to be fermionic.   By a bosonic state we mean a state on which
the GSO operator $(-1)^F$ has eigenvalue $+1$, while on a
fermionic state it has eigenvalue $-1$.  We claim that such a
system describes a collection of $p$ D-branes and $q$
D-antibranes.

To verify this, we examine the spectrum, considering first Type II
superstrings (so that the open strings are oriented). The D-D
strings are states both of whose ends have bosonic Chan-Paton
labels.  $(-1)^F$ acts by $+1$ on the Chan-Paton wavefunction of
such a string, so for a state to have $(-1)^F=1$ overall, the
internal part of the wavefunction (made from the string
oscillators) must also have $(-1)^F=1$.  This leads to the
standard GSO projection, giving a spectrum with a massless gauge
field and no tachyon.  Likewise, a $\bar {\rm D}$-$\bar {\rm D}$
string has a fermionic label at each end, so $(-1)^F$ acts on the
Chan-Paton wavefunction by $(-1)^2=+1$, and again the $\bar {\rm
D}$-$\bar {\rm D}$ strings have the conventional GSO projection.
But for   D-$\bar{\rm D}$ strings, the result is different.  In
this case, there is a bosonic label at one end of the string and a
fermionic one at the other end, so the Chan-Paton wavefunction has
$(-1)^F=-1$.  Hence to get an overall value $(-1)^F=1$, the
internal wavefunction must likewise have $(-1)^F=-1$. The
projection onto these states is the opposite of the usual GSO
projection, and it gives for the D-$\bar{\rm D}$ strings a
spectrum with a tachyon and no massless gauge field, and opposite
chirality for massless fermions in the Ramond (R) sector.  These are
of course the standard results for Type II.

We now move on to the case of Type I superstrings with $p$ D9 and
$q$ D${\bar 9}$ branes.  Again we represent this system by
allowing $p$ bosonic and $q$ fermionic labels for the Chan-Paton
factors.  (For tadpole cancellation, one eventually wants
$p-q=32$.)  The novel ingredient is that the open strings are
unoriented; we must define an operator $\Omega$ that exchanges the
two ends of an open string, and project onto states with
$\Omega=1$.   For D9-D9 strings in Type I, the $\Omega$ projector
is the usual one that leaves an $SO(p)$ gauge group.  (If we take
the opposite projector, we get instead the theory discussed in
section 3.4 below.) For D9-D$\bar 9$ and D$\bar 9$-D9 strings, the
$\Omega$ projector simply identifies the D9-D${\bar 9}$ and D$\bar
9$-D9 states, so the correct choice of $\Omega$ projector is
needed to find the correct wavefunctions but not to find the
spectrum.  What about the D$\bar 9$-D$\bar 9$ strings?  In this
case, we want to show that the $\Omega$ projector is the usual one
in the Neveu-Schwarz (NS) sector, but is opposite to the usual one in
the R sector.  If that is true,  the Chan-Paton wavefunctions
for massless D$\bar 9$-D$\bar 9$ strings in the NS
sector will be the usual antisymmetric tensors
$\psi_{ij}=-\psi_{ji}$, $i,j=1,\dots,q$, leading to massless
$SO(q)$ gauge fields.  But the Chan-Paton wave functions of
massless  D$\bar 9$-D$\bar 9$ strings in the R sector will be
symmetric tensors $\chi_{ij}=\chi_{ji}$, which is the spectrum of
left-handed massless MW fermions from D$\bar 9$-D$\bar 9$ strings
that was claimed above.

To reduce the question to a standard one, note that defining the
correct $\Omega$ operator is equivalent to knowing how to compute
the Moebius strip contribution to the open string partition
function. This is for a familiar reason: a Moebius strip
worldsheet can arise from a trace in the open string Hilbert space
with a factor of $\Omega$ inserted. Now the Moebius strip has only
a single boundary.  Let us represent the Moebius strip in the
standard fashion as the strip $0\leq \sigma^1\leq \pi$,
$-\infty<\sigma^2<\infty$ in the $\sigma^1-\sigma^2$ plane, with
the equivalence
\begin{equation}
\sigma^1\to \pi-\sigma^1,~~\sigma^2\to \sigma^2+2\pi t.
\end{equation}
The Moebius strip has only one boundary component, a circle $C$
that we can identify as $\sigma^1=0$, $0\leq \sigma^2\leq 4\pi t$.
A path integral on the Moebius strip computes $\Tr\,(-1)^{\beta
F}\Omega e^{-2\pi   t H}$, where $\beta\in{0,1}$ determines the
spin structure in the $\sigma^2$ direction and $F$ is the number
of worldsheet fermions.

If the Chan-Paton labels are bosonic, then $p$ Chan-Paton states
propagating around $C$ give a factor of $p$ in the path integral,
multiplying the usual evaluation of $\Tr\,(-1)^{\beta F}\Omega
e^{-2\pi tH}$ by $p$ and leaving unchanged the usual $\Omega$
projection. But $q$ fermionic Chan-Paton states propagating around
$C$ give a factor of $+q$ if the spin structure restricted to $C$
is in the NS sector, and $-q$ if the spin structure
restricted to $C$ is in the R sector. \footnote{In other
words, we get $+q$ if worldsheet fermions restricted to $C$ are
antiperiodic and $-q$ if they are periodic.  In the former case,
the path integral has a factor which is a trace in the Chan-Paton
Hilbert space, to which all states contribute $+1$, while in the
latter case the trace is replaced by $\Tr\,(-1)^F$, and fermionic
states contribute $-1$.}  If the Chan-Paton states contribute a
factor of $q$ to $\Tr\,(-1)^{\beta F}\Omega e^{-2\pi tH}$, this
corresponds to the usual $\Omega$ projection, but if they
contribute a factor $-q$, we interpret this in an operator
language to mean that the sign of $\Omega$ is reversed, and we get
the opposite-to-usual $\Omega$ projection.

The path integral on the Moebius strip, viewed in the open string
channel as computing $\Tr\,(-1)^F\Omega e^{-2\pi H}$, receives
contributions from both the NS and R sectors of
the open string. A standard result, explained on p. 42 of
\cite{Polchinski:1998}, is that the spin structure on $C$ is
actually the same as that of the open strings. Thus, looking at
the Moebius strip in the closed string channel, a closed string
wrapped once around $C$ is in the NS-NS sector if the open strings
are in the NS sector, and in the R-R sector if the open strings are
in the R sector. (In particular, the spin structure on $C$ does
not depend on whether the worldsheet fermions are periodic or
antiperiodic under $\sigma^1\to \pi-\sigma^1,\,\sigma^2\to
\sigma^2+2\pi t$.  That is because $C$ covers the range $0\leq
\sigma^2\leq 4\pi t$ and so wraps twice around the $\sigma^2$
direction.)

Applied to our problem, the statement about the spin structure on
$C$ means that, as we have claimed above, the open strings in the
NS sector have the standard $\Omega$ projection, and those in the
R sector have the opposite projection. For completeness, we
summarize how the claim about the spin structure on $C$ is
established. The transformation $\sigma^1\to \pi-\sigma^1$,
$\sigma^2\to \sigma^2+2\pi t$, exchanges left- and right-moving
worldsheet fermions $\psi $ and $\tilde \psi$, so under this
transformation
\begin{equation}\label{tranform}
\psi(\sigma^1,\sigma^2+2\pi t)=-(-1)^\beta\tilde\psi(\pi-\sigma^1,\sigma^2),
\end{equation}
where $\beta\in\{0,1\}$ determines the spin structure of the
M\"obius strip  in the $\sigma^2$ direction.  It is convenient to
combine $\psi$ and $\tilde \psi$ to a function defined for all
values of $\sigma^1$; we extend $\psi$ to $\sigma^1<0$ by setting
$\psi(-\sigma^1,\sigma^2)=\tilde\psi(\sigma^1, \sigma^2)$, and to
$\sigma^1>\pi$ by setting $\psi(\sigma^1,\sigma^2) =
(-1)^\alpha\psi(\pi-\sigma^1,\sigma^2)$, where $\alpha\in {0,1}$,
with $\alpha=0$ for the R sector of open strings and
$\alpha=1$ for the NS sector. Altogether
$\psi(\sigma^1,\sigma^2)$ is naturally extended to a field
periodic in $\sigma^1$ with
\begin{equation}\label{natex}
\psi(\sigma^1+2\pi,\sigma^2)=(-1)^\alpha\psi(\sigma^1,\sigma^2)
\end{equation}
In terms of the extended field $\psi$, (\ref{tranform}) can be
written
\begin{equation}
\psi(\sigma^1,\sigma^2+2\pi t) =-(-1)^\beta\psi(\sigma^1-\pi,\sigma^2).
\end{equation}
By applying this relation twice, we get
\begin{equation}
\psi(\sigma^1,\sigma^2+4\pi t)= \psi(\sigma^1-2\pi,\sigma^2)=(-1)^\alpha
\psi(\sigma^1,\sigma^2).
\end{equation}
So the worldsheet fermions at $\sigma^1=0$ transform under $\sigma^2
\to \sigma^2+4\pi$ by $(-1)^\alpha$, and hence the spin structure on
$C$ is the same as the spin structure in the open string channel at
$\sigma^2=0$, as was to be shown.

\subsection{Anomaly Analysis}

Having identified the spectrum, we can now analyze the anomalies.
We have already arranged for the cancellation of the ${\rm tr}
R^6$ term, but there is quite a bit more that needs to be checked.
The formulas must reduce to the standard ones of the type I theory
for $n = 0$, so it is really the $n$ dependent terms that are in
issue. We will let the index 1 refer to the $SO(32+n)$ group and
the index 2 refer to the $SO(n)$ group.  By standard
manipulations, one finds that just as in the $n=0$ case the total
anomaly 12-form factorizes into a product of a four-form and an
eight-form, $I_{12}= 2Y_4 Y_8$, where
\begin{equation}
Y_4 = {1 \over 2} \big( {\rm tr} R^2 - {\rm tr} F_1^2 + {\rm tr}
F_2^2 \big)
\end{equation}
and
\begin{equation}
Y_8 =  {1 \over 24} \Big( {1 \over 8}{\rm tr} R^4 + {1 \over 32}
({\rm tr} R^2 )^2 + ({\rm tr} F_1^4 - {\rm tr} F_2^4) -{1 \over 8}
{\rm tr} R^2 ({\rm tr} F_1^2 - {\rm tr} F_2^2) \Big) ,
\end{equation}
where the subscript 1 refers to $SO(n+32)$ and the subscript 2
refers to $SO(n)$.
This factorization assures us that that there is a Chern--Simons
counterterm for which the anomaly cancels, just as in the $n=0$
case.

There is a slick derivation of these results using techniques
pioneered in \cite{Morales:1999ux} \cite{Scrucca:1999jq}. 
(See also \cite{Stefanski:1999yx} concerning the anomalous couplings
of D-branes and O-planes.) The $99$
open strings give left-handed Majorana--Weyl fermions in the
antisymmetric tensor representation (A) of $SO(32+n)$, the
$\bar9\bar9$ open strings give left-handed Majorana--Weyl fermions
in the symmetric tensor representation (S) of $SO(n)$, and the
$9\bar9$ open strings give right-handed Majorana--Weyl fermions in
the bifundamental representation. Thus the anomaly polynomial
$I_{12}$ is proportional to the 12-form piece of
\begin{equation} \label{expression1}
{1 \over 2}\Big( {\rm Tr}_A e^{iF_1} + {\rm Tr}_S e^{iF_2} - {\rm tr}
e^{iF_1}{\rm tr} e^{iF_2}\Big) \hat A(R) +{1\over 16} L(R).
\end{equation}
The last term is the contribution of the gravitino and the
dilatino from the closed-string sector. It is expressed here as
half of the contribution of a self-dual R-R field.  ($L(R)$ is the
Hirzebruch L-polynomial.) The reason this is correct is that the
anomaly cancellation of the Type IIB theory implies that the
contribution of a self-dual R-R field cancels that of {\it two}
Majorana--Weyl gravitinos and dilatinos.

Using the identities
\begin{equation}
{\rm Tr}_A e^{iF} = {1 \over 2} \Big([{\rm tr} e^{iF}]^2 - {\rm
tr} e^{2iF} \Big)
\end{equation}
and
\begin{equation}
{\rm Tr}_S e^{iF} = {1 \over 2} \Big([{\rm tr} e^{iF}]^2 + {\rm
tr} e^{2iF} \Big),
\end{equation}
the expression (\ref{expression1}) can be recast in the form
\begin{equation}
{1 \over 4} \Big([{\rm Str} e^{iF}]^2 - {\rm Str} e^{2iF} \Big)
\hat A(R) +{1\over 16} L(R),
\end{equation}
where ${\rm Str} e^{iF} =  {\rm tr} e^{iF_1} - {\rm tr} e^{iF_2}$.
Remarkably, the 12-form piece of this expression agrees
with the 12-form piece of ${1\over 4}Y^2$, where $Y$ is given by
\begin{equation} \label{Yformula}
Y = {\rm Str} e^{iF} \sqrt{ \hat A(R) } - 32 \sqrt{L(R/4)}.
\end{equation}
The key to proving this is the identity \cite{Morales:1999ux}
\cite{Scrucca:1999jq}
\begin{equation}
\sqrt{ \hat A(R) L(R/4) } = \hat A(R/2).
\end{equation}
This identity is an immediate consequence of the defining
relations
\begin{equation}
 \hat A(R) = \prod_i {\lambda_i/2 \over {\rm sinh} \lambda_i /2}
\end{equation}
and
\begin{equation}
  L(R) = \prod_i {\lambda_i \over {\rm tanh} \lambda_i }.
\end{equation}
In writing the anomaly 12-form as $Y^2/4$, one also needs to know
that the 12-form part of ${\rm Str}e^{iF}\hat A(R/2)$ is the
same as the 12-form part of $2^{-6}{\rm Str}e^{2iF}\hat A(R)$, since
that 12-form is homogeneous of degree 6 in $F$ and $R$.

Thus, just as in the IIB case, the anomaly $I_{12}$ is
proportional to the 12-form part of $I=Y^2$. Since we take the net
D9-brane charge to be 32, the 0-form part of $Y=Y_0+Y_4+Y_8+\dots$
vanishes, so the characteristic classes $Y_4$ and $Y_8$ are the
two leading terms in the expansion of $Y$, and $I_{12}$ is
proportional to $Y_4Y_8$.

The first term in the expression (\ref{Yformula}) for $Y$ can be
interpreted as the D-brane (or world-sheet boundary) contribution,
whereas the second term is the orientifold plane (or cross-cap)
contribution. The zero-form piece of $Y$ cancels between the two
terms. This is the cancellation of the R-R tadpole.  The three
terms in $Y^2$ correspond to the contributions of the annulus, the
Moebius strip, and the Klein bottle.

The original study of the type I anomaly utilized the R-R 2-form
$C_2$ but not the dual R-R 6-form $C_6$. In order to give a
symmetrical (but informal) treatment, analogous to that described
for the IIB theory in the previous section, we will include both
fields in the discussion that follows. Accordingly we define $C =
C_2 - C_6$. This enables us to proceed exactly as in the IIB case.
Specifically, we again introduce
\begin{equation}
S_{\rm CS} = \mu \int C Y \, ,
\end{equation}
where $\mu$ is a normalization constant.  Then we define $\Omega$
by $Y = d \Omega$ and denote the variation of $\Omega$ under local
gauge and Lorentz transformations by $\delta \Omega = d \Lambda$.
Since only the four-form and eight-form pieces of $Y$ contribute,
only the two-form and six-form pieces of $\Lambda$ are relevant.
It then follows that the anomaly cancels for
\begin{equation}
\delta C = \Lambda,
\end{equation}
and that the gauge-invariant R-R field strengths are given by $H =
dC - \Omega$.

It is natural to suppose that the tachyon dependence could be
added in the same way as in the type IIB theory. Specifically, in
the formula for $Y$ we would replace ${\rm tr} \, e^{iF_1}- {\rm
tr}\, e^{iF_2}$ by ${\rm Str} e^{i{\cal F}}$, where now the
curvature of the superconnection is the $(2n + 32) \times (2n +
32)$ matrix
\begin{equation}
i {\cal F}  = \left( \begin{array}{cc}i F_1 - T  T^t & D T^t
\\ DT & i F_2 - T^t T \end{array} \right) ,
\end{equation}
since $T$ is real.  The relation to brane annihilation is the same
as before.

\subsection{Theories with Symplectic Groups}

It was noted in \cite{Sugimoto:1999tx} that it is possible to
construct a tachyon-free nonsupersymmetric theory by modifying the
orientifold projection that is used in constructing the Type I
theory out of the Type IIB theory. Specifically, instead of
modding out by $\Omega$ -- the usual $Z_2$ symmetry, discussed in
section 3.2, that is utilized in the Type I construction -- one
mods out by the $Z_2$ symmetry $\Omega' = \Omega (-1)^{F_L}$,
where $F_L$ is the world-sheet fermion number for left-movers. The
resulting theory contains an orientifold plane which has R-R
charge $+32$, which is the opposite sign from the usual case.
Accordingly, it is necessary to add $32$ anti-D9-branes to cancel
the R-R charge. Moreover, in this situation the gauge group
associated to the anti-D9-branes is the symplectic group that is
variously called $Sp(16)$ or $USp(32)$.

This theory is tachyon-free, but it contains a tree-level dilaton
potential term. The orientifold plane has positive tension (as
well as positive charge). Thus, in contrast to the usual Type I
theory, this tension reinforces the tension of the D9-branes, to
give a total vacuum energy of $64 T_{D9}$ \cite{Sugimoto:1999tx}.
In string metric this includes a factor of $e^{-\phi}$, where
$\phi$ represents the dilaton. In canonical metric this becomes
$e^{3\phi/2}$. Therefore, the runaway dilaton would appear to
drive the system to zero coupling.  It may be possible to achieve
stability and finite coupling with  a warped compactification
\cite{Dudas:2000ff} \cite{Blumenhagen:2000dc}. At any rate, a
weak-coupling anomaly analysis in ten-dimensional Minkowski space should
make sense.

The closed string spectrum (at zero coupling) is identical to that
of the Type I theory. In particular, it contains a massless
gravitino. This is somewhat surprising, since the open string
sector associated to the anti-D9-branes is not supersymmetric. In
addition to the $Sp(16)$ gauge bosons, the massless spectrum
contains Majorana-Weyl fermions in the antisymmetric tensor
representation of the gauge group. In contrast to the case of
orthogonal groups, this multiplet is not the adjoint, which is a
symmetric tensor. In fact, it is reducible into a part that has no
symplectic trace and a singlet. The singlet can be identified as
the Goldstino associated with the broken supersymmetry. Both the
gravitino and the Goldstino have only one chirality, a situation
that would be impossible in a maximally symmetric ten-dimensional
spacetime (such as Minkowski or de Sitter space), for
supersymmetry breaking in such a spacetime would require the
gravitino to get mass, and this is only possible if both
chiralities are present. Supersymmetric invariance of a theory
with such an {\it inedible Goldstino}, which has only one
chirality and cannot combine with the graviton to give a massive
state, depends on the dilaton potential term in the effective
action, as was demonstrated in \cite{Dudas:2000nv}.  Because this
term is present, such a theory does not lead to a maximally
symmetric ten-dimensional spacetime, and the chirality of the
gravitino and Goldstino leads to no contradiction.

Even though this theory has many differences from the Type I
theory, the anomaly analysis works in exactly the same way. The
antisymmetric tensor representation of $Sp(16)$ has dimension 496,
just as in the $SO(32)$ case. Moreover, all of its charges with
respect to the maximal torus (Cartan subalgebra) are identical to
those in the $SO(32)$ case. Therefore, since the massless closed
string sector is identical to that of the Type I theory, the
anomaly analysis is the same. Moreover, as discussed in
\cite{Sugimoto:1999tx}, one can also add additional brane
antibrane pairs to make an unstable theory with $Sp(16+n) \times
Sp(n)$ gauge symmetry. The anomaly analysis of this system is
identical to that of the $SO(32+2n) \times SO(2n)$ theory
described in the previous section.

\section{Another Class of Ten-Dimensional Models}

Three examples of ten-dimensional tachyon-free string theories
without spacetime supersymmetry are known. The first one
discovered is the $SO(16)\times SO(16)$ heterotic theory
\cite{Dixon:1986iz} \cite{Alvarez-Gaume:1986jb}. Since heterotic
theories do not have D-branes, this example does not lend itself
to the type of analysis we are doing here. A second example is the
$Sp(16)$ theory discussed in the previous section. The third
example, which is the subject of this section, has $U(32)$ gauge
symmetry. It was discovered by Sagnotti in \cite{Sagnotti:1995ga}
\cite{Sagnotti:1997qj}. (For a review and a discussion of related
models in lower dimensions see \cite{Sagnotti:2000xd}.) We will
quickly recall the essential features of this theory and then
consider including additional brane antibrane pairs to give a
theory with $U(32+n) \times U(n)$ gauge symmetry. Other examples
of tachyon-free models without supersymmetry, which we will not
consider here, have been constructed in lower dimensions
\cite{Antoniadis:1999xk} \cite{Aldazabal:1999jr}.

\subsection{The U(32) Theory}

A variant of the Type IIB superstring theory, usually called the
Type 0B theory, is constructed by making a different GSO
projection from the usual one \cite{Dixon:1986iz}
\cite{Seiberg:1986by}. It also gives a modular invariant partition
function and therefore is perturbatively consistent. The Type 0B
theory is a theory of oriented closed strings only. It involves no
orientifold plane or spacetime filling D-branes. The spectrum
includes the closed string tachyon, which is ordinarily removed by
the GSO projection. The entire spectrum contains bosons only,
since the R-NS and NS-R sectors are both projected out. At the
massless level, the NS-NS spectrum is the same as for Type II
superstrings: a graviton, two-form, and dilaton. The massless R-R
spectrum is double that of the IIB theory. By this we mean that
there is no self-duality constraint on the R-R spectrum.
Evidently, the 0B theory is nonchiral, and so it is trivially
anomaly-free.

Sagnotti's $U(32)$ theory is constructed as an orientifold
projection of the Type 0B theory. The resulting theory has
unoriented breakable strings, rather like the Type I theory. The
orientifold projection removes the tachyon and half the massless
R-R fields from the spectrum. The remaining massless R-R fields
are identical to those of the IIB theory. The massless NS-NS
spectrum consists only of the graviton and the dilaton, just as in
the Type I theory. There are still no fermions in the closed
string spectrum. Thus the closed string spectrum includes just one
chiral field: the four-form R-R potential whose field strength is
self-dual. Clearly, if this were the whole story, the theory would
be anomalous. However, there is an orientifold plane that carries
$-32$ units of R-R charge (just as in the Type I theory) which
requires the addition of 32 spacetime-filling D9-branes.
Properties of some of the other D-branes in this theory have been
discussed in \cite{Blumenhagen:1999ns} \cite{Dudas:2000sn}.

As was explained by Sagnotti, and will be clear from the anomaly
analysis, the massless open-string spectrum contains $U(32)$ gauge
fields. It also contains Weyl fermions that belong to the
antisymmetric tensor representation (496) of the gauge group.
Sagnotti showed that these fermions together with the self-dual
R-R field give anomalies that can be cancelled by the addition of
a Chern-Simons term in the usual way. It is an amusing fact that
whereas the cancellation of the ${\rm tr} R^6$ term in the Type I
theory anomaly polynomial requires 496 Majorana--Weyl fermions to
cancel the contributions of the gravitino and dilatino, the
cancellation in this case takes place between the contributions of
the self-dual R-R field and 496 Weyl fermions. It is clear that
this is the right counting, since we know that in the Type IIB
theory the gravitational anomaly contributions of the self-dual
R-R field precisely cancel those of a pair of Majorana--Weyl
gravitinos and dilatinos.

\subsection{Addition of Brane-Antibrane Pairs}

Let us now consider adding $n$ D9-branes and $n$ anti-D9-branes to
the $U(32)$ theory. Clearly, this will give a theory that has
$U(32+n) \times U(n)$ gauge symmetry. Also, it contains complex
tachyon fields in the bifundamental representation. Like the Type
IIB and Type I theories with added brane-antibrane pairs, which
were analyzed in previous sections, we expect this theory to be
perturbatively well-behaved. Therefore the anomalies should
cancel.

It is easy to figure out what the chiral fermions are in this
case. The $99$ open strings give left-handed Weyl fermions in the
antisymmetric tensor representation of $U(32+n)$, the $\bar9\bar9$
open strings give left-handed Weyl fermions in the symmetric
tensor representation of $U(n)$, and the $9\bar9$ open strings
give right-handed Weyl fermions in the bifundamental
representation. The anomalies can be computed as usual, and one
finds that the anomaly polynomial $I_{12}$ is proportional to the
12-form piece of
\begin{equation} \label{expression}
\Big( {\rm Tr}_A e^{iF_1} + {\rm Tr}_S e^{iF_2} - {\rm tr}
e^{iF_1}{\rm tr} e^{iF_2}\Big) \hat A(R) +{1\over 8} L(R),
\end{equation}
where the subscript 1 refers to $U(n+32)$ and the subscript 2
refers to $U(n)$. The last term is the contribution of the
self-dual R-R field. This expression looks just like
(\ref{expression1}), though now we are dealing with unitary groups
instead of orthogonal groups. Just as in the previous case, the
expression (\ref{expression}) can be recast in the form
\begin{equation}
{1 \over 2} \Big([{\rm Str} e^{iF}]^2 - {\rm Str} e^{2iF} \Big)
\hat A(R) +{1\over 8} L(R),
\end{equation}
where ${\rm Str} e^{iF} =  {\rm tr} e^{iF_1} - {\rm tr} e^{iF_2}$.
Again, the 12-form piece of this expression agrees with the
12-form piece of ${1\over 2}Y^2$, where $Y$ is given by
\begin{equation}
Y = {\rm Str} e^{iF} \sqrt{ \hat A(R) } - 32 \sqrt{L(R/4)}.
\end{equation}

One difference from the $SO(32+n) \times SO(n)$ Type I theory is
that traces of odd powers of $F$ are now nonzero. The anomaly can
now be cancelled by the addition of a Chern--Simons term of the
form $\mu \int C Y$. This formula is more like that of the IIB
case in that $C = -i C_0 + C_2 + i C_4 - C_6 -i C_8$, with the
same self-duality constraint as in that case. This example
provides a pleasing mix of features of the Type I and Type IIB
theories. Note that the relative $U(1)$ gauge field eats the R-R
scalar, just as in the Type IIB problem. The dependence on
bifundamental tachyons could be added in the same way that it was
added there.

\section{Type I Compactified on a Smooth K3}

When the Type I superstring theory is compactified on a smooth K3,
the requirement that $dH = {\rm tr} R^2 - {\rm tr } F^2$ should be
exact implies that there are 24 units of instanton number. This
can be achieved by a combination of large instantons embedded in
the SO(32) gauge group and small instantons (which are D5-branes)
localized on the K3 \cite{Witten:1996gx}. If there are $k$
coincident D5-branes, the associated world-volume theory is an
$Sp(k)$ gauge theory. In this case, the remaining $24-k$ units of
instanton number must be embedded in the $SO(32)$ gauge group,
which breaks it to $SO(8+k)$.

To be specific, we will focus on the case $k=24$, which gives the
maximal symmetry group $SO(32) \times Sp(24)$. This theory has 32
D9-branes and 24 coincident D5-branes. The resulting 6d theory has
${\cal N} =1$ supersymmetry (8 conserved supercharges). The
massless fields consist of the supergravity multiplet, one tensor
multiplet, vector multiplets for each of the gauge groups, and
three classes of hypermultiplets. There are 20 hypermultiplets of
gravitational origin, which are gauge singlets. The second class
of hypermultiplets arises as zero modes of 55 strings. They belong
to the antisymmetric tensor representation of the $Sp(24)$ gauge
group. This representation is reducible because it contains a
symplectic traceless part and a singlet. The singlet provides the
requisite Goldstone fermion. The third class of hypermultiplets
arises as zero modes of 59 strings. They belong to the
bifundamental representation, of course. These states actually
belong to ``half hypermultiplets," which is possible because the
representation is pseudoreal.

The anomaly analysis of these systems and the other compactified
systems we will consider could be analyzed from either a
ten-dimensional or a six-dimensional viewpoint. In the former case
the fivebranes would be described as localized defects embedded in
ten dimensions and anomalies would be analyzed locally taking
account of the phenomenon of ``anomaly inflow.'' In the
alternative six-dimensional viewpoint, one simply considers the
effective six-dimensional theory that arises at length scales
large compared to the compactification scale. This is the approach
we will take in the following discussion. For example, in the
particular example under consideration at the moment, altogether
there are 244 more hypermultiplets than vector multiplets. This is
the number required for the cancellation of the ${\rm tr} R^4$
piece of the anomaly 8-form in ${\cal N} =1$ 6d theories with one
tensor multiplet \cite{Green:1985bx}.

The anomaly 8-form for this 6d theory was analyzed in detail in
\cite{Schwarz:1996zw}, where it was shown that it factorizes in
the form
\begin{equation}
I_8 = - {1\over 16} ({\rm tr} R^2 - {\rm tr} F_9^2) ( {\rm tr} R^2
+ 2 {\rm tr} F_9^2 - 2 {\rm tr} F_5^2).
\end{equation}
Here $F_9$ refers to the $SO(32)$ group and $F_5$ refers to the
$Sp(24)$ group. It follows that the gauge and Lorentz anomalies
can be cancelled by adding a Chern--Simons term of the form $\int
C_2 Y_4$ and requiring that $C_2$ transform under Lorentz and
gauge transformations in the usual fashion. Our goal in the
remainder of this section is to explore how this analysis should
be generalized when one allows for the addition of anti-D9-branes
and anti-D5-branes.

We add $n$ D9-brane anti-D9-brane pairs and $m$ D5-brane
anti-D5-brane pairs. Then the D9-branes and anti-D9-branes give
the gauge group
\begin{equation}
G_9 = SO(32+n) \times SO(n)
\end{equation}
just as we found for the uncompactified theory in the preceding
section. Taking all the D5-branes and all the anti-D5-branes to be
coincident\footnote{It does not matter whether the D5-branes are
coincident with the anti-D5-branes, since any massless fermions
arising in this way would be nonchiral. This has to be the case,
because the two sets of branes can be separated without changing
the gauge groups.} gives the gauge group
\begin{equation}
G_5 = Sp(24+m) \times Sp(m) .
\end{equation}
This theory is unstable and nonsupersymmetric with tachyons, like
the examples described in the preceding sections. As was done for
those examples, we will identify the various chiral fermions,
compute the associated anomaly polynomial, and deduce the
Chern-Simons term of the effective theory.

Let us now consider the massless chiral fermions. We use the
convention that fermions that have the same chirality as the
supercharge (and hence of fermions in vector supermultiplets) are
left-handed and contribute to the anomaly polynomial with a plus
sign. Right-handed fermions, such as the ones in hypermultiplets,
contribute with a minus sign. Here and in section 6, handedness of
fermions is understood in the six-dimensional sense.

Let us start with the zero modes of strings connecting the various
9-branes. Since the 99 spectrum is supersymmetric, these fermions
are left-handed and belong to the adjoint representation. The
$9\bar9$ fermions are right-handed (as follows, for example, from
the discussion in section 3.2) and belong to the bifundamental
representation. The $\bar9 \bar9$ fermion zero modes are
left-handed and belong to the symmetric representation of $SO(n)$,
just as we found previously in ten dimensions. Thus the Chern
characters associated with these states give
\begin{equation} {\rm Tr} e^{i F_9} - {\rm tr} e^{i F_9}\,
{\rm tr} e^{i F_{\bar9}} + {\rm Tr}_S e^{i F_{\bar9}} = {1\over
2}\, ({\rm Str} e^{i F_9})^2 - {1\over 2}\,{\rm Str} e^{2i F_9}.
\end{equation}
As usual, Tr refers to the adjoint representation, tr to the
fundamental representation, and ${\rm Tr}_S$ to the symmetric
tensor representation.

Next let us consider the 55 spectrum. As before, the 55 strings
give left-handed fermions in the adjoint representation and
right-handed fermions in the antisymmetric tensor representation
of the $Sp(24+m)$ gauge group. For the $\bar5 \bar5$ strings the
chiralities are reversed: the adjoint fermions are right-handed
and the antisymmetric tensor representation fermions are
left-handed.\footnote{A $\bar5$-brane differs from a $5$-brane by
reflection of one of its worldvolume coordinates together with one
of the normal coordinates, so the $\bar5 \bar5$ spectrum can be
deduced directly from the $55$ spectrum.  Alternatively, according
to the reasoning in section 3.2, going from 55 to $\bar5 \bar5$
should reverse the $\Omega$ projection in the R sector,
exchanging the roles of symmetric and antisymmetric tensors of
$Sp(m)$ or in other words exchanging the adjoint representation
with the antisymmetric tensor.} As explained in a footnote above,
there is no contribution from $5 \bar5$ strings. Altogether these
states contribute
\begin{equation} {\rm Tr} e^{i F_5} - {\rm Tr} e^{i
F_{\bar5}} - {\rm Tr}_A e^{i F_5} + {\rm Tr}_A e^{i F_{\bar5}} =
{\rm Str} e^{2i F_5}.
\end{equation}

Finally, we have four classes of fermion zero modes for 59
strings. In each case we get a bifundamental representation, with
the factor of $1/2$ explained earlier. The 95 and $\bar9\bar5$
fermion zero modes are right-handed whereas the $9\bar5$ and
$\bar9 5$ ones are left-handed. Thus these contributions give
\begin{equation} -{1 \over 2}\, {\rm tr} e^{i F_9}\, {\rm tr} e^{i F_5} -{1
\over 2}\, {\rm tr} e^{i F_{\bar9}}\, {\rm tr} e^{i F_{\bar5}} +{1
\over 2}\, {\rm tr} e^{i F_{\bar9}}\, {\rm tr} e^{i F_5}+{1 \over
2}\, {\rm tr} e^{i F_9}\, {\rm tr} e^{i F_{\bar5}} = -{1 \over
2}\, {\rm Str} e^{i F_9}\, {\rm Str} e^{i F_5}
\end{equation}

Adding up the three sets of terms given above and expanding gives
\begin{equation}
-224 + 6 {\rm Str} F_5^2 - 3 {\rm Str} F_9^2 + {1 \over 8} \Big(
{\rm Str} F_9^2\Big)^2 -{1\over 8} {\rm Str} F_9^2 {\rm Str} F_5^2
+\ldots
\end{equation}
Including the gravitational contributions, one finds that the
anomaly 8-form factorizes
\begin{equation}
I_8 = - {1\over 16}Y_4^{(1)} Y_4^{(2)},
\end{equation}
where
\begin{equation}
Y_4^{(1)} = {\rm tr} R^2 - {\rm Str} F_9^2
\end{equation}
and
\begin{equation}
Y_4^{(2)} =  {\rm tr} R^2 + 2 {\rm Str} F_9^2 - 2 {\rm Str} F_5^2.
\end{equation}
Given the result in the absence of the extra brane-antibrane
pairs, this is the simplest outcome we could have expected.

It follows that the anomaly cancellation works as before.
For example, if we define $Y_4^{(1)} = d \Omega_3$ and $\delta
\Omega_3 = d \Lambda_2$, then a Chern--Simons term of the form
$\mu \int C_2 Y_4^{(2)}$ can cancel the anomaly provided that we
require $\delta C_2 = \Lambda_2$. As usual, $H= dC_2 - \Omega_3$
is then gauge invariant.

In a more symmetrical treatment $ C_2 Y_4^{(2)}$
would be replaced by $C_2^+(Y_4^{(2)} + Y_4^{(1)}) +
C_2^-(Y_4^{(2)} - Y_4^{(1)}) $, where $C_2^{\pm} = {1\over 2} (C_2 \pm
\tilde C_2)$. The self-dual part $C_2^+$ belongs to the supergravity multiplet,
and the anti-self-dual part $C_2^-$ belongs to the tensor supermultiplet.
(This separation is difficult to achieve covariantly, but that is a problem
we face in all the examples with self-dual R-R forms.)

\section{Type I Compactified on a $T^4/Z_2$ Orbifold}

\subsection{Review of the Basic Model}

The study of a particular compactification of the Type I theory on
a $T^4/Z_2$ orbifold was explored in \cite{Gimon:1996rq}. The
anomaly cancellation analysis of this model was subsequently
carried out in \cite{Berkooz:1996iz}. This orbifold is a singular
limit of a K3, so once again it is necessary to account for 24
units of instanton number. In the model of \cite{Gimon:1996rq} one
unit of instanton number is localized at each of the 16 fixed
points of the orbifold and the remaining 8 units are realized as
D5-branes. Equivalently, we can look at the R-R charge. Each
orbifold point has a charge of $-3/2$ arising from the geometry
and $+1$ from the embedded instanton for a total of $-1/2$. Even
though none of the instantons is embedded in the $SO(32)$ gauge
group associated with the D9-branes (except as point instantons
that do not break the gauge symmetry), the ${\bf Z_2}$ that is
used to form the orbifold action acts on the gauge group breaking
it to $U(16)$. This group can be broken further by Wilson lines,
as we will discuss.

Altogether the model has a gauge group of the form $G_5 \times
G_9$, where $G_5$ is associated with the D5-branes and $G_9$ is
associated with the D9-branes. There is a rich set of
possibilities for each. Those for the D5-branes can be understood
geometrically, whereas those for the D9-branes correspond to
various possibilities for the Wilson lines. Remarkably the two
stories are isomorphic, and the full set of models that can be
realized this way is invariant under T-duality. In other words,
for every construction there is a dual construction for which the
role of the D9-branes and the D5-branes is interchanged.

Let us recall the geometric description of the possible gauge
groups $G_5$. Just as in the preceding section, when $k$ D5-branes
coincide at a regular point of the the orbifold, the associated
world-volume theory is an ${\cal N} =1$ $Sp(k)$ gauge theory.
However, when fivebranes approach an orbifold point, the mirror
images come into play and the group is enhanced to $U(2k)$. Thus,
for example, if all eight of them are on the same orbifold point
this gives $G_5 =U(16)$. Another interesting fact is that half
fivebranes can attach to orbifold points, so that unitary groups
of odd rank are also possible. Only an integral number of
fivebranes can move off of an orbifold point. One extreme case is
to attach a half fivebrane to each of the orbifold points giving
the group $G_5 = U(1)^{16}$. This configuration is special, in
that it is the only one that cancels the R-R charge locally.
However, it is not necessary to do that. Thus there are a number
of topologically distinct sectors characterized by the number (and
locations) of orbifold points that have a half fivebrane attached
to them. As required by T duality, there are corresponding
statements that could be made about the possible configurations of
Wilson lines and their implications for $G_9$. According to
\cite{Berkooz:1996iz}, most of the topological sectors suffer a
nonperturbative anomaly that makes them unacceptable, even though
they are all consistent perturbatively. The rule is that the
number of orbifold points with attached half fivebranes must be
either 0, 8, or 16. Since one also has the corresponding
restriction for the Wilson lines, there are 9 different
topological sectors that are allowed, but only 6 of them are
distinct when T duality is taken into account.

We will review how the anomaly analysis works for this class of
models. Later we will generalize the results to allow for the
addition of anti-D5-branes and anti-D9-branes. For describing the
analysis it is convenient to assign all the fivebranes to orbifold
points, so that $G_5 = \prod_1^{16} U(m_i)$, where $\sum_1^{16}
m_i =16$. Note that we include all the orbifold points in the sum,
since we allow the $m_i$'s to be zero. This helps to keep the
notation relatively simple. Situations in which there are
fivebranes in the bulk correspond to a simple Higgsing of this
class of models. On the other hand, the anomaly analysis of these
models is more subtle than those with only fivebranes in the bulk,
because of the $U(1)$ factors. Therefore by considering the model
with this choice of $G_5$ and $G_9 = \prod_1^{16} U(n_i)$, where
$\sum_1^{16} n_i =16$, we are really taking account of all
interesting cases.

\subsection{The Massless Spectrum}

As in the preceding section, the massless gauge singlet fields,
associated to zero modes of closed strings, consist of the
supergravity multiplet, a tensor multiplet, and 20
hypermultiplets. One significant fact is that 16 of these
hypermultiplets are associated to the 16 orbifold points. One of
the scalars in each of the hypermultiplets belongs to the R-R
sector. The corresponding 16 R-R scalar fields play an important
role in the anomaly analysis.

Let us now consider the part of the massless spectrum that arises
as zero modes of open strings. Most of it is pretty obvious.
Consider the $5_i 5_j$ open string spectrum first. It is clear
that there is no contribution from open strings connecting
fivebranes at one orbifold point to ones at a different orbifold
point, since they are spatially separated. So we need only
consider the case $i=j$. These give vector supermultiplets for
each of the $U(m_i)$ groups. In addition, they also give two
hypermultiplets belonging to antisymmetric tensor representations.
The spectrum of 99 strings is completely analogous, as required by
T duality.

We now turn to the spectrum of zero modes of $5_i 9_j$ open
strings. This spectrum, which was analyzed in
\cite{Berkooz:1996iz}, turns out to be quite subtle. It is pretty
evident that one expects each of the possible 59 open strings to
give a massless hypermultiplet in the bifundamental
representation. The subtlety is that there are two possibilities
$(m_i, n_j)$ and $(m_i, \bar n_j)$.\footnote{Recall that a
hypermultiplet contains the complex conjugate, so $(m_i, n_j)$ is
the same as $(\bar m_i, \bar n_j)$ and $(m_i, \bar n_j)$ is the
same as $(\bar m_i, n_j)$.}  For any given pair $i,j$, the
distinction between these two cases is a matter of convention, as
it can be reversed by complex conjugation of one of the groups.
However, in general, there are more distinct pairings than there
are unitary groups, so this distinction is not entirely
convention. Therefore let us distinguish the two possibilities by
a parameter $w_{ij}$ that takes the value $+1$ for values of $i$
and $j$  such that the 59 open strings transform as $(m_i,n_j)$
and the value $-1$ in the second case. This matrix of parameters
satisfies certain properties that we will describe later.

\subsection{Anomaly Analysis}

Let us now focus on the Chern characters that enter in the anomaly
analysis. The $9_i 9_i$ chiral fermions give
\begin{equation}
{\rm Tr}e^{iF_{9i}} - 2 {\rm Tr}_A e^{iF_{9i}} = n_i - 2 {\rm tr}
F_{9i}^2 + 2 \Big( {\rm tr} F_{9i} \Big)^2 +{2 \over 3} {\rm tr}
F_{9i}^4 -{2\over 3} {\rm tr} F_{9i} {\rm tr} F_{9i}^3 + \dots
\end{equation}
Similarly, the $5_i 5_i$ chiral fermions give
\begin{equation}
{\rm Tr}e^{iF_{5i}} - 2 {\rm Tr}_A e^{iF_{5i}} = m_i - 2 {\rm tr}
F_{5i}^2 + 2 \Big( {\rm tr} F_{5i} \Big)^2 +{2 \over 3} {\rm tr}
F_{5i}^4 -{2\over 3} {\rm tr} F_{5i} {\rm tr} F_{5i}^3 + \dots
\end{equation}
The $5_i 9_j$ chiral fermions, with the rule described above, give
\[
-{\rm tr} e^{iF_{5i}} {\rm tr} e^{i w_{ij} F_{9j}} = - m_i n_j +
{1 \over 2} m_i {\rm tr} F_{9j}^2 + {1 \over 2} n_j {\rm tr}
F_{5i}^2 + w_{ij} {\rm tr} F_{5i} {\rm tr} F_{9j}
\]
\begin{equation}
- {1 \over 24} m_i {\rm tr} F_{9j}^4 - {1 \over 24} n_j {\rm tr}
F_{5i}^4 -{1\over 4} {\rm tr} F_{5i}^2 {\rm tr} F_{9j}^2 -{1\over
6} w_{ij} {\rm tr} F_{5i}^3 {\rm tr} F_{9j} - {1 \over 6} w_{ij}
{\rm tr} F_{5i} {\rm tr} F_{9j}^3 + \ldots
\end{equation}

Combining all of the above give
\[
- 224 + 6 \sum_i {\rm tr} F_{9i}^2 +  6 \sum_i {\rm tr} F_{5i}^2
-{ 1 \over 4} \sum_i {\rm tr} F_{5i}^2 \sum_j {\rm tr} F_{9j}^2 \]
\[ + 2 \sum_i \Big( {\rm tr} F_{9i} \Big)^2 + 2 \sum_i \Big( {\rm
tr} F_{5i} \Big)^2 + \sum_{ij} w_{ij} {\rm tr} F_{5i} {\rm tr}
F_{9j}\] \[ -{2 \over 3} \sum_i {\rm tr} F_{9i} {\rm tr} F_{9i}^3
-{2 \over 3} \sum_i {\rm tr} F_{5i} {\rm tr} F_{5i}^3 -{1 \over 6}
\sum_{ij} w_{ij} {\rm tr} F_{5i}^3 {\rm tr} F_{9j}  -{1 \over 6}
\sum_{ij} w_{ij} {\rm tr} F_{5i} {\rm tr} F_{9j}^3.
\]
Multiplying this by $\hat A(R)$, adding the closed-string
contribution, and extracting the 8-form gives
\[
I_8 = -{ 1 \over 16} \Big( {\rm tr} R^2 \Big)^2 + { 1\over 8} {\rm
tr} R^2 \Big( \sum_i {\rm tr} F_{9i}^2 + \sum_i {\rm tr} F_{5i}^2
\Big) -{1 \over 4} \sum_i {\rm tr} F_{5i}^2  \sum_j {\rm tr}
F_{9j}^2 \] \[ + { 1\over 24} {\rm tr} R^2 \Big( \sum_i ({\rm tr}
F_{9i})^2 + \sum_i ({\rm tr} F_{5i})^2  + {1 \over 2} \sum_{ij}
w_{ij} {\rm tr} F_{5i} {\rm tr} F_{9j}\Big)\] \[ -{2 \over 3}
\sum_i {\rm tr} F_{9i} {\rm tr} F_{9i}^3 -{2 \over 3} \sum_i {\rm
tr} F_{5i} {\rm tr} F_{5i}^3  -{1 \over 6} \sum_{ij} w_{ij} {\rm
tr} F_{5i}^3 {\rm tr} F_{9j}  -{1 \over 6} \sum_{ij} w_{ij} {\rm
tr} F_{5i} {\rm tr} F_{9j}^3. \] The key identity
\cite{Berkooz:1996iz} that makes it possible to factorize this in
the way required to achieve anomaly cancellation is
\begin{equation}
\sum_{i=1}^{16} w_{ij} w_{ik} = 16 \delta_{jk}.
\end{equation}
Using this, we can recast $I_8$ in the factorized form
\begin{equation} \label{I8}
I_8 = -{1 \over 4} Y_4^{(5)} Y_4^{(9)}  -{2 \over
3}\sum_{i=1}^{16} Y_{2i} Y_{6i},
\end{equation}
where
\begin{equation}
Y_4^{(5)} = {1\over 2} {\rm tr} R^2 -\sum_i {\rm tr} F_{5i}^2
\end{equation}
\begin{equation}
Y_4^{(9)} = {1\over 2} {\rm tr} R^2 -\sum_i {\rm tr} F_{9i}^2
\end{equation}
\begin{equation}
Y_{2i} = {\rm tr} F_{5i} + {1 \over 4} \sum_j w_{ij}{\rm tr}
F_{9j}
\end{equation}
\begin{equation}
Y_{6i} = {\rm tr} F_{5i}^3 + {1 \over 4} \sum_j w_{ij}{\rm tr}
F_{9j}^3 -{1 \over 16} {\rm tr} R^2 Y_{2i}.
\end{equation}
Thus anomaly cancellation is achieved by adding Chern--Simons
terms of the form \[ \mu \int ( C_2 Y_4^{(9)} + \sum_{i=1}^{16}
C_{0i} Y_{6i} ).\] Here, $C_{0i}$ are the 16 R-R scalars that were
pointed out earlier. This completes the review of results from
\cite{Berkooz:1996iz}. Now, we are ready to consider adding
additional brane-antibrane pairs.

\subsection{Addition of Brane-Antibrane Pairs}

Let us now consider adding additional brane-antibrane pairs. As
before we will only consider the case when all the D5-branes and
anti-D5-branes are located at the orbifold points so that the
gauge group is a product of unitary groups. Letting $m_i$ denote
the number of half-D5-branes and $\tilde m_i$ the number of half
anti-D5-branes at the $i$th orbifold point, the gauge group is
\begin{equation}
G_5 = \prod_{i=1}^{16} U(m_i)\times  U(\tilde m_i).
\end{equation}
Each of the $m_i$ and $\tilde m_i$ is allowed to be either zero or
a positive integer. Since the total fivebrane charge must be 8,
the only perturbative restriction is
\begin{equation}
\sum_{i=1}^{16} \Big( m_i - \tilde m_i \Big) = 16.
\end{equation}
The D9-branes are treated in similar manner, so that
\begin{equation}
G_9 = \prod_{i=1}^{16} U(n_i)\times  U(\tilde n_i),
\end{equation}
where
\begin{equation}
\sum_{i=1}^{16} \Big( n_i - \tilde n_i \Big) = 16.
\end{equation}

Let us now consider the spectrum of chiral fermions arising from
open-string zero modes \cite{Aldazabal:1999jr}. As before the $9_i
9_i$ and $5_i 5_i$ open strings give left-handed fermions in the
adjoint representation and two copies of right-handed fermions in
the antisymmetric tensor representation. The $\bar9_i \bar9_i$ and
$\bar5_i \bar5_i$ open strings give left-handed fermions in the
adjoint representation and two copies of right-handed fermions in
the {\it symmetric} tensor representation. The $9_i 5_j$ and
$\bar9_i \bar5_j$ open strings give right-handed fermions in a
bifundamental representation, whereas the $9_i \bar5_j$ and
$\bar9_i 5_j$ opens strings give left-handed fermions in a
bifundamental representation. In each of these four cases the
issue of whether one has (fundamental, fundamental) or
(fundamental, antifundamental) is described by the parameters
$w_{ij}$ introduced earlier.

At this point, it is easy to verify that the number of
right-handed fermions exceeds the number of left-handed ones by
244, as required. However, this is not yet the whole story. There
are still $9_i \bar9_i$ and $5_i \bar5_i$ open-string zero modes
to be taken into account. They do provide additional chiral
content, even though they give an equal number of left-handed and
right-handed fermions. The correct rule in each case is that they
give (fundamental, fundamental) and (antifundamental,
antifundamental) left-handed fermions and (fundamental,
antifundamental) and (antifundamental, fundamental) right-handed
fermions. The contribution of this set of states to the Chern
character term in the anomaly is
\begin{equation}
\Big( {\rm tr} e^{iF_{9i}} - {\rm tr} e^{- iF_{9i}}\Big) \Big(
{\rm tr} e^{iF_{\bar9i}}- {\rm tr} e^{-iF_{\bar9i}}\Big) = -4 {\rm
tr} \, {\rm sin} F_{9i} {\rm tr} \, {\rm sin} F_{\bar9i}.
\end{equation}

Taking all of the above into account, it is straightforward to
recompute the anomaly 8-form $I_8$. The answer turns out to be
just what one might have guessed. Namely, $I_8$ is still given by
eq. (\ref{I8}), where now
\begin{equation}
Y_4^{(5)} = {1\over 2} {\rm tr} R^2 -\sum_i {\rm Str} F_{5i}^2
\end{equation}
\begin{equation}
Y_4^{(9)} = {1\over 2} {\rm tr} R^2 -\sum_i {\rm Str} F_{9i}^2
\end{equation}
\begin{equation}
Y_{2i} = {\rm Str} F_{5i} + {1 \over 4} \sum_j w_{ij}{\rm Str}
F_{9j}
\end{equation}
\begin{equation}
Y_{6i} = {\rm Str} F_{5i}^3 + {1 \over 4} \sum_j w_{ij}{\rm Str}
F_{9j}^3 -{1 \over 16} {\rm tr} R^2 Y_{2i}.
\end{equation}
Thus the anomaly cancellation works as before with the
substitution of supertraces for traces. This seems to be the
general rule.

\section*{Acknowledgments}
Part of this material was presented in the lecture given by JHS at the
``Davidfest'' Conference held at the ITP in Santa Barbara on March 2--3, 2001.
This work was supported in part by the U.S. Dept. of Energy under
Grant No. DE-FG03-92-ER40701 and by the Caltech Discovery Fund.

\end{document}